\documentstyle[epsf]{mn}

\def\MN{{ MNRAS, }}

\def\ApJS{{ ApJS, }}
\def\AsA{{ A\&A, }}

\def\AJ{{ AJ, }}
\def\ARAA{{ ARAA,}}

\title[Cepheid Proper Motions]
{Galactic Kinematics of Cepheids from Hipparcos Proper Motions
\thanks{Based on data from the Hipparcos astrometry satellite}}
\author[M. W. Feast and P. A. Whitelock ]
{Michael Feast $^{1}$ and Patricia Whitelock $^{2}$  \\
$^{1}$ Astronomy Department, University of Cape Town, 7700 Rondebosch,
South Africa.\\
email: mwf@uctvax.uct.ac.za \\
$^{2}$ South African Astronomical Observatory, PO Box 9, 7935 Observatory,
South Africa. \\
email: paw@saao.ac.za \\}

\begin{document}

\maketitle

\begin{abstract}
 The Hipparcos proper motions of 220 galactic Cepheids, together with
relevant ground-based photometry, have been analysed. The effects of
galactic rotation are very clearly seen. Mean values of the Oort constants,
$A= 14.82 \pm 0.84\ {\rm km\ s^{-1}kpc^{-1}}$, and $B = -12.37 \pm 0.64\ {\rm
km\ s^{-1}kpc^{-1}}$, and of the angular velocity of circular rotation at the
Sun, $\Omega_{\rm o}= 27.19 \pm 0.87\ {\rm km\ s^{-1}kpc^{-1}}$, are
derived. Comparison of the value of $A$ with values derived from recent
radial velocity solutions confirm, within the errors, the zero-points of the
PL and PLC relations derived directly from the Hipparcos trigonometrical
parallaxes of the same stars. The proper motion results suggest that the
galactic rotation curve is declining slowly at the solar distance from the
galactic centre ($\rm (d \Theta/dR)_{o} = -2.4 \pm 1.2\ km\
s^{-1}kpc^{-1}$). The component of the solar motion towards the North
Galactic Pole is found to be $ + 7.61 \pm 0.64\ \rm km\ s^{-1}$. Based on
the increased distance scale deduced in the present paper the distance to
the galactic centre derived in a previous radial velocity study is increased
to $\rm R_{o}= 8.5 \pm 0.5\ kpc$.
 \end{abstract}


\section{Introduction}
 The measurement of trigonometrical parallaxes and proper motions of Cepheid
variables by the Hipparcos satellite (ESA 1997) represents a major advance
over earlier work because of the scope and accuracy of the new results and
the fact that they are referred to a co-ordinate system which is uniform
over the whole sky and based on the positions of distant extragalactic
objects. The link between the Hipparcos reference frame and the
International Celestial Reference System (ICRS) based on the positions of
extragalactic radio sources implies that the Hipparcos proper motions are
quasi-inertial to within $\rm \pm 0.25\ mas\ yr^{-1}$ (ESA 1997 volume 1
section~1.2). In a previous paper (Feast \& Catchpole 1997 = paper~1), the
zero-point of the Cepheid period-luminosity (PL) relation was derived from
Hipparcos trigonometrical parallaxes. The present paper is primarily
concerned with an analysis of the Hipparcos proper motions. We also use the
Hipparcos trigonometrical parallaxes to derive zero-points for PL and
period-luminosity-colour (PLC) relations which have been recently adopted in
discussions of the radial velocities of Galactic Cepheids. This allows us to
revise Galactic constants derived from the radial velocity studies. The
constants of absolute and differential rotation of our Galaxy are derived
from the Hipparcos proper motions. The value of the Oort constant of
differential rotation, $A$, derived from the proper motions is essentially
independent of the adopted distance scale; whilst that derived from radial
velocities is nearly inversely proportional to this scale. Thus a comparison
allows us to confirm the Cepheid luminosity scale derived from the
trigonometrical parallaxes.

\section{Observational Data}
   The Cepheids discussed in the present paper are listed in Table~1. This
table gives the Hipparcos Catalogue number (HIP), the variable star name and the
values of $\langle V \rangle $ and ${\langle B \rangle -\langle V \rangle }$ adopted. These latter values are from the
sources discussed in paper~1. The table also gives the log of the period or,
for the stars noted in paper~1 as pulsating in the first overtone, the log
of the fundamental period calculated in the manner described there.  These
latter stars are denoted by ``o" in the last column of the table. The
effects of overtone, and possible overtone, pulsators on the trigonometrical
parallax work were discussed in detail in paper~1 so far as the 26 stars
used for the adopted PL zero-point were concerned. Any likely effect on the
PL zero-point of an uncertainty with regard to pulsation mode was shown to
be very small and this also applies to the trigonometrical parallax analyses
of the present paper. It remains possible that there are a limited number of
unrecognised overtone pulsators amongst the full set of stars listed in
Table~1. Most of our results from the proper motions (e.g.\ concerning $A$
and $\Omega_{o}$) are quite insensitive to distance errors and thus to a
possible misidentification of the pulsation mode of some stars (see, e.g.\
sections 5 and 6 below).  Second-order terms and values of the local solar
motion will be affected by any such misidentification but the effect is
likely to be very small. For instance the value of $w_{o}$ (defined in
section 9 below) might be slightly underestimated due to this cause.

In carrying out the analyses two slightly different sets of reddenings
($E_{B-V}$) have been used. When a PL relation has been used to estimate
distances the reddenings were derived from the period-colour relation used
in paper~1, viz;
 \begin{equation}
B-V = 0.416 \log P + 0.314
\end{equation}          
(Laney \& Stobie 1994).
 The advantage of doing this is discussed in paper~1. However, this advantage
no longer applies when the distances are being derived from a PLC relation.
In that case the reddenings listed in Table~1 were used. These were taken
where possible from the compilation of Caldwell \& Coulson (1987) and are
on the two-colour, $BVI$, system with a zero-point derived from Cepheids in
open clusters. It should be noted that reddenings in the $BVI$ system
provide the basis for the PC relation (equation 1). For stars which are not in
the Caldwell \& Coulson compilation, reddenings were taken from Fernie
(1990) or from determinations on the same system (Fernie et al.\ 1995,
electronic catalogue). With the adopted reddenings, values of $A_{V}$
were derived using equation 7 of paper~1 (see Laney \& Stobie 1993).

The astrometric data can be obtained directly from the Hipparcos Catalogue
(ESA 1997). This gives the trigonometrical parallaxes of the stars together
with their standard errors. The parallaxes of the 26 Cepheids of highest
weight in the PL zero-point solution of paper~1 are given in that paper. The
Hipparcos Catalogue also gives for each star the proper motions in Right
Ascension and Declination ($ \mu_{\alpha *} = \mu_{\alpha} \cos \delta$ and
$ \mu_{\delta}$), their standard errors and the coefficient of correlation
between the two proper motion components. All the astrometric data are
expressed in milli-arcsec (mas).

\section{PLC and PL zero-points}
 When in later sections of this paper the results from the Hipparcos proper
motions are compared with radial velocity studies we shall be referring
principally to two recent radial velocity studies of Cepheid kinematics,
viz., Metzger et al.\ (1997 =~MCS) and Pont et al.\ (1994 = PMB). In deriving
distances MCS use a PL relation whilst PMB use a PLC relation. The Hipparcos
trigonometrical parallaxes are used below to derive zero-points of these
relations. A comparison with the zero-points actually used by MCS and PMB
then indicates the scaling factor necessary to bring their results onto the
scale set by the trigonometrical parallaxes.
 
MCS use a PL relation of the form:
\begin{equation}
\langle M_{V} \rangle  = -2.87 \log P + \rho_{1}
\end{equation}
 with $\rho_{1} = -1.23$. Revised values of $\rho_{1}$ have been derived
using precisely the same methods and data as used to derive the PL relation
in paper~1 and for the same groupings of stars. Four solutions are given in
Table~2. They correspond to solution~A and numbers 1, 2, 4 and 6 of paper~1
Table~2. In paper~1 additional solutions (solutions B) were given taking
into account possible errors in addition to those of the parallaxes. Since
these were shown there to be only marginally different from solutions A they
are not given in the present case. As with the analogous solution~in paper~1
we adopt solution~4 as the best value of $\rho_{1}$ ($ = -1.38 \pm 0.09$).
Thus the Hipparcos parallaxes indicate that the MCS zero-point should be
brightened by 0.15 mag or their distance scale increased by 7 percent. It
may be noted that the interstellar reddenings adopted by MCS are on the
$BVI$ system of Caldwell \& Coulson (1987). As with those adopted in paper~1
they have as their zero-point the reddenings of Cepheids in open clusters.
Thus these two systems of reddening should be closely similar.

PMB use a PLC relation of the form:
\begin{equation}
\langle M_{V} \rangle  = -3.80\log P + 2.70(\langle B \rangle -\langle V \rangle )_0 + \rho_{2}
\end{equation}
 (see Feast \& Walker 1987).  PMB adopt $\rho_{2} = -2.27$. In deriving a
value of $\rho_{2}$ from the Hipparcos trigonometrical parallaxes, the
general procedure of paper~1 was again followed.  But for the reasons given
in section~2 (above), the reddenings listed in Table~1 were used. For the 26
stars of highest weight (which are the same 26 stars listed in Table~1 of
paper~1) all the reddenings come directly from Caldwell \& Coulson (1987).

There should be negligible intrinsic scatter in the PLC relation. Martin et
al.\ (1979) (see also Feast \& Walker 1987) found that a sample of LMC
Cepheids showed a scatter of $0.14 \pm 0.02$ about a PLC relation. This
scatter includes observational error, errors in reddening corrections and
distance scatter within the LMC. It should therefore be an upper limit to
the scatter in the sample of stars used with the Hipparcos parallaxes,
especially since these are brighter and generally more extensively observed
photometrically.  Note that because the coefficient of the colour term in
the PLC relation is similar to the ratio $A_{V}/E_{B-V}$, the PLC relation
is relatively insensitive to errors in the reddenings. Following the method
discussed in paper~1, the PLC zero-point was derived taking into account the
errors in the parallaxes and the possible scatter in the PLC relation. Table
3 shows the results with this latter scatter ($\sigma _{PLC}$) taken as 0,
0.10, or 0.20 mag. The solutions correspond to the four solutions of Table
2. It is clear that within reasonable limits the value of $\sigma _{PLC}$
has little effect. We adopt solution~4 with $\sigma _{PLC} = 0.10$ ($-2.38
\pm 0.10$). This is 0.11 mag brighter than than the zero-point used by PMB,
or an increase in the distance scale of 5 percent.

\section{The Adopted Proper Motions}
 The Hipparcos proper motions and their errors (see section~2 above) were
converted into components in galactic longitude and latitude ($ \mu_{\ell *}
= \mu_{\ell} \cos b$, and $ \mu_{b}$) and their errors, using the constants
provided in the Hipparcos Catalogue and taking into account the correlation
between proper motions in Right Ascension and Declination. Then if $\mu$ is
the proper motion in mas and $d$ is the distance in kpc, the corresponding
velocity is $\kappa \mu d\ {\rm km\ s^{-1}}$ where $\kappa = 4.74047$ (a value
also taken from the Hipparcos Catalogue).

\begin{figure*}
\centering
\epsffile[-7 85 425 340]{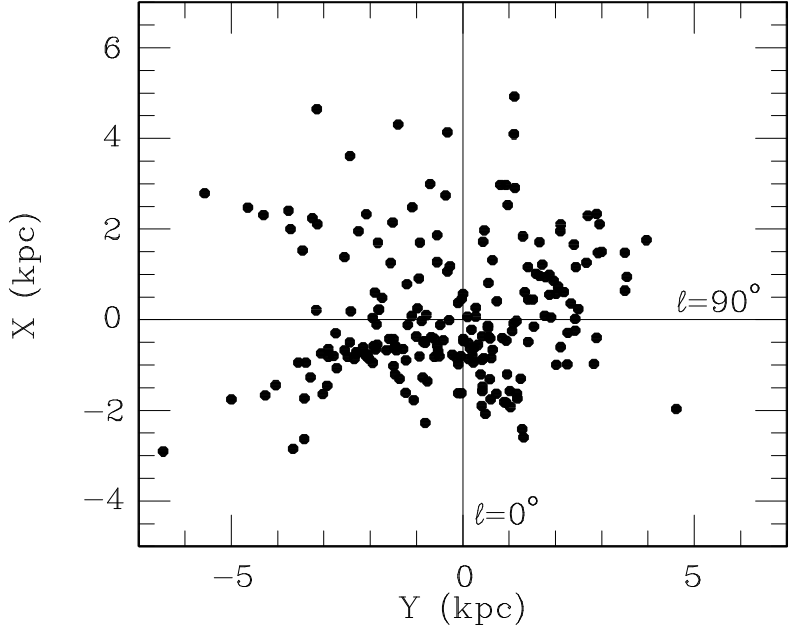}
 \caption{ The distribution of the Cepheids used in the proper motion
solutions seen projected onto the galactic plane.  The Sun is at the origin
of the co-ordinate system. The distances are from the PL relation derived in
paper~1.}
 \end{figure*}

In analysing the proper motions account must be taken in the weighting not
only of the standard errors of the proper motion components themselves
($\sigma_{\ell *}, \sigma_{b}$) but also the scatter about an adopted
galactic model due to the random motions of the stars. The weights were
taken as proportional to the reciprocals of the squares of $\sigma_{\kappa
\mu \ell *}$ and $\sigma_{\kappa \mu b}$, where,
 \begin{equation}
\sigma_{\kappa \mu \ell *}^{2} = (\kappa \sigma_{\ell *})^{2} +
(\alpha^{2} \sin^{2} l + \beta^{2} \cos^{2} l)/d^{2}
\end{equation}
and 
\begin{eqnarray}
\sigma_{\kappa \mu b}^{2} & = &  (\kappa \sigma_{b})^{2} + 
(\alpha^{2} \cos^{2} \ell \sin^{2} b  \nonumber \\
&&+ \beta^{2} \sin^{2} \ell \sin^{2} b + \gamma^{2} \cos^{2} b)/d^{2} .
\end{eqnarray}
 $\rm \alpha, \beta$ and $\rm \gamma$ are the axes of the velocity ellipsoid
for which we have adopted the values 13, 9 and 5 $\rm km\ s^{-1}$ for
Cepheids from Delhaye (1965)(see also PMB).

Of the Cepheids for which astrometric data were available from Hipparcos,
the following stars have been omitted in the analyses for the reasons given
below. In most cases the stars were rejected because of the large peculiar
velocities (given below) derived in preliminary analyses using PL distances
and the proper motions. H10, H29 and H30 indicate an anomalous entry in the
relevant field, Hn, of the Hipparcos catalogue (for more details see the
explanatory supplement). In the following H10 indicates that the star is
flagged in H10 as being in a binary or multiple system; H29 indicates that
20 percent or more of the data were rejected in the astrometric solution and
H30 indicates that the goodness of fit statistic is greater than 3, possibly
implying a poor astrometric solution.\\
 DP Vel; $\rm HIP\, 46610$; Photometry too sparse; see paper~1.\\
AW Per; $\rm HIP\, 22275$; Binary; see paper~1.\\
AX Cir; $\rm HIP\, 72773$; Binary; see paper~1; H10, H30.\\
UX Per; $\rm HIP\, 10332$; Peculiar velocity $\rm \sim 500\ km\ s^{-1}$; H10, H29, H30.\\
SS CMa; $\rm HIP\, 36088$; Peculiar velocity $\rm \sim 200\ km\ s^{-1}$.\\
V Vel:  $\rm HIP\, 45949$; Rejected by PMB due to discrepant radial velocity.
Rejected by Caldwell \& Coulson (1987) due to discrepant colours.
Peculiar velocity $\rm \sim 80\ km\ s^{-1}$.\\
SU Cru; $\rm HIP\, 59996$; Rejected by PMB due to
discrepant radial velocity. Peculiar velocity $\rm \sim 240\ km\ s^{-1}$;
H10, H29.\\
SY Nor; $\rm HIP\, 77913$; Peculiar velocity $\rm \sim 100\ km\ s^{-1}$; H10, H30.\\
TW Nor; $\rm HIP\, 78771$; Peculiar velocity $\rm \sim 100\ km\ s^{-1}$. This is an important
star in, or in the direction of, the open cluster Lynga 6. Caldwell \&
Coulson (1987) rejected it as a PL calibrator because it gave a discrepant
zero-point but more recently Laney \& Stobie (1994) have included it in
a zero-point determination by the cluster method.

\begin{figure*}
\centering
\epsffile[7 85 425 340]{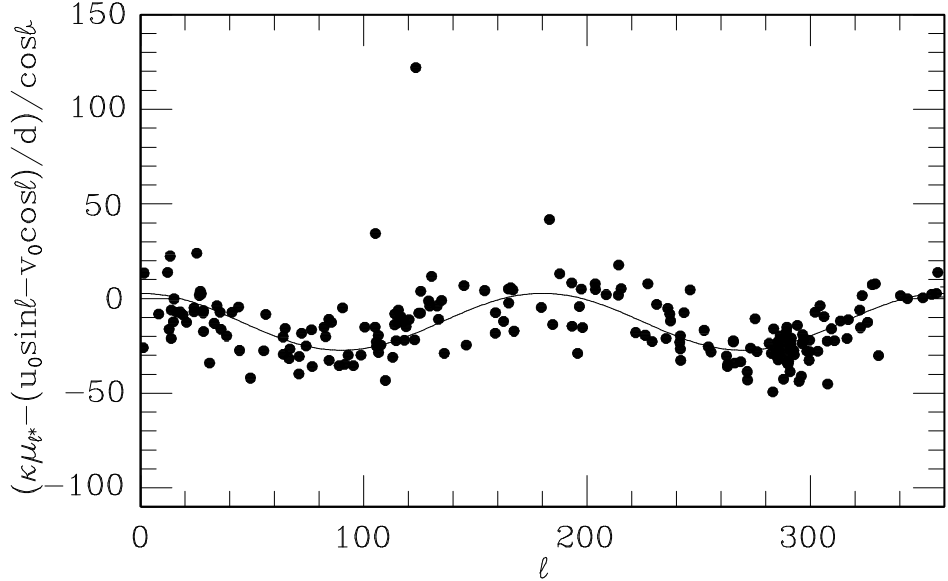}
 \caption{The proper motion in galactic longitude multiplied by $\kappa$,
and corrected for local solar motion, plotted against galactic longitude. The
curve corresponds to solution~2 of Table~4. The three outstanding stars are
nearby and have low weight in the solution (see text for further details).}
 \end{figure*}

\section{First Order Solution for Galactic Rotation}
 It is useful to begin with a simple, first order (``Oort" type) solution
for galactic rotation from the proper motions. The relevant equation is
generally written in the form;
 \begin{equation}
\kappa \mu_{\ell *} = (u_{\rm o} \sin \ell -v_{\rm o} \cos \ell)/d + (A \cos 
2l + B) \cos b ,
\end{equation} 
 where $u_{\rm o}$ and $v_{\rm o}$ are the components of the local solar
motion (with respect to young objects) towards the galactic centre and in
the direction of galactic rotation. The Oort constants are defined by
 \begin{equation}
A = - \frac{1}{2} R_{\rm o}(d\Omega/dR)_{\rm o} 
\end{equation}
and 
\begin{equation}
B = -\Omega_{\rm o} - \frac{1}{2} R_{\rm o}(d\Omega/dR)_{\rm o},
\end{equation}
 where $R$ is the distance of the star from the galactic centre and $\rm
\Omega$ the angular velocity (=$\rm \Theta /R$, $\rm \Theta$ being the
circular velocity). Subscript ``o'' indicates the values of these quantities
at the solar position. However, the quantity of interest is often not $B$
itself but
 \begin{equation}
A - B = \Omega_{\rm o} = \Theta _{\rm o}/ R_{\rm o}.
\end{equation}
 It is useful therefore to solve equation 6 in the form:
 \begin{equation}
\kappa \mu_{\ell *} = (u_{\rm o} \sin \ell - v_{\rm o} \cos \ell)/d + (2A 
\cos^{2} \ell - \Omega _{\rm o}) \cos b.
\end{equation}
 Thus here and in the rest of the paper we work in terms of $A$ and
$\Omega_{\rm o}$, but explicit values of $B$ are given for our final
solutions. Because of the uniformity of the Hipparcos reference frame and
the fact that it is tied to the positions of extragalactic objects, the
values of $B$ and $ \Omega_{\rm o}$ (which are directly affected by any
rotation of the reference frame) should be considerably more reliable than
previous values.

There have been a large number of determinations from radial velocities of
the components of the local solar motion from Cepheids.  These stars are
expected in the mean to be in circular rotation about the Galactic Centre. 
There is some discussion of the local solar motion, relative to extreme
population~I objects, in section~9, below. We
adopt in this section either $u_{\rm o} = 9$  and $v_{\rm o} = 13 \ \rm km\
s^{-1}$ or $u_{\rm o} =9.32$  and $ v_{\rm o} = 11.18\ {\rm km\ s^{-1}}$.
The latter  is from PMB. Table~4 shows a number of solutions for $A$ and
$\Omega_{\rm o}$. Except in the case of solution~3, the distances used in
these solutions were obtained with the PL relation derived in paper~1 from
the Hipparcos trigonometrical parallaxes, viz.
 \begin{equation}
\langle M_{V} \rangle  = -2.81 \log P - 1.43,
\end{equation}
 together with reddenings obtained from equation 1 above. For solution~3 the
``PLC" distances discussed in section~3 were used.  Solution~4 shows the
effect of decreasing all distance moduli by 0.4 mag. In solutions 5 and 6
the data are divided into two  groups on the basis of the stellar
distance, and in solution~7 the effect of forcing $2A$ to equal $\Omega_{\rm
o}$, as is required for a flat rotation curve ($\rm \Theta = constant$), is
shown.

These various solutions show that the derived constants are rather
insensitive to the adopted distance scale. There is a suggestion that the
derived value of $A$ increases with distance (though the increase is within
the errors). This may be due both to local streaming motions and (for the
more distant stars) to the neglect of higher-order terms in the equation
used (equation 10). Forcing a flat rotation curve on the data leads to a
distinctly smaller value of $A$ ($13.34 \pm 0.42$) than that derived from
the other solutions.

Figure 1 shows a plot of the distances of Cepheids with Hipparcos proper
motions projected onto the Galactic Plane (PL distances). Whilst many
previous studies of galactic rotation from proper motions have referred to a
relatively small region around the sun, the Hipparcos data covers a
significant region of the Galactic Disc.

The quantity $(\kappa \mu_{\ell *} -(u_{\rm o} \sin \ell - v_{\rm o}\cos
\ell)/d)/\cos b$ is shown in Fig 2 plotted against galactic longitude
($\ell$) for solution~2  of Table~4. Figures 3 and 4 show similar plots
for the nearer and more distant stars (solutions 5 and 6), the curves in
each case  illustrate the corresponding solutions in Table~4.

\begin{figure*}
\centering
\epsffile[7 85 425 340]{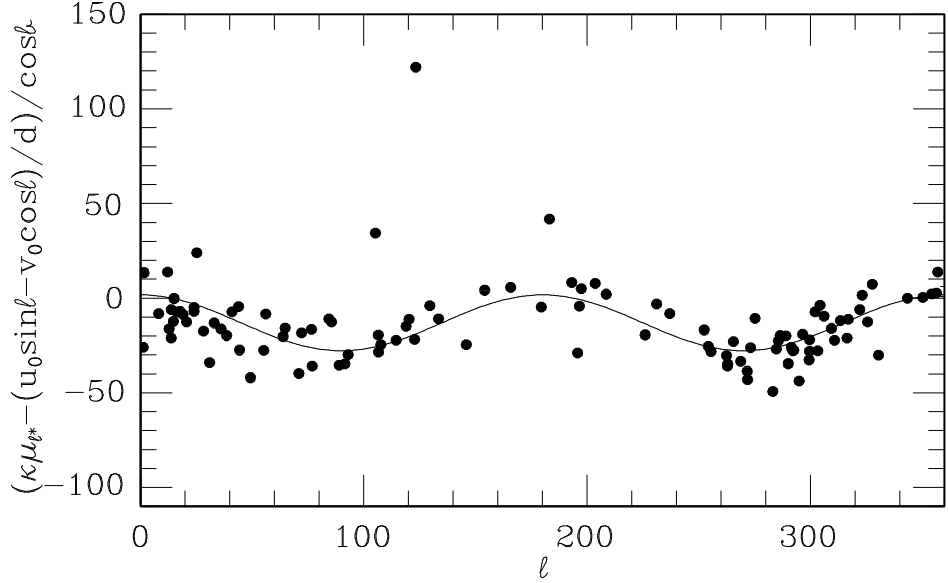}
\caption{ As Fig 2 but for stars with distances less than 2 kpc. The
curve  illustrates  solution~5 of Table~4.}
\end{figure*}

\begin{figure*}
\centering
\epsffile[7 85 425 340]{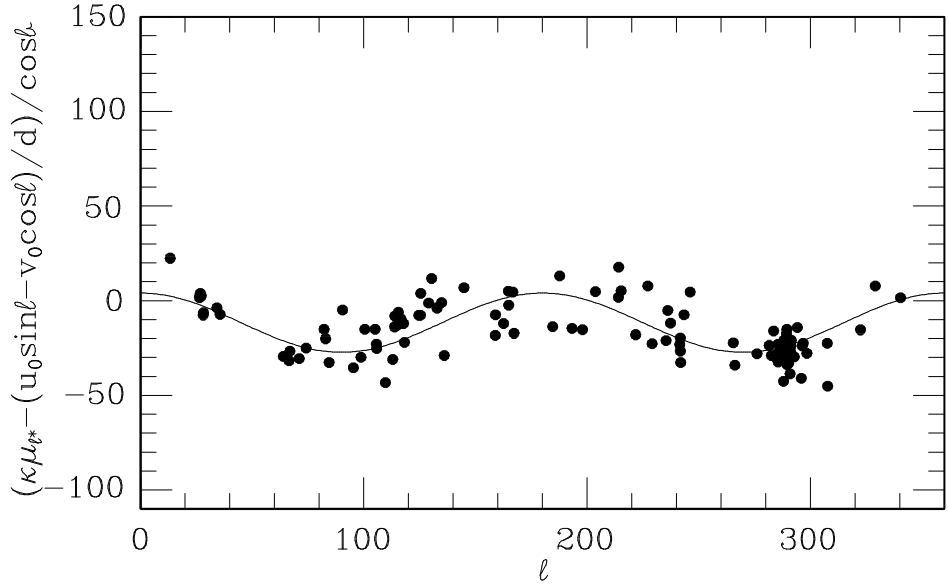}
\caption{As Fig 2 but for stars with distances equal to  or greater than 2
kpc.
The curve  illustrates solution~6 of Table~4.}
\end{figure*}

Whilst the effect of Galactic rotation on proper motions has long been known
(e.g.\ Oort 1927), it does not seem to have been possible prior to the
Hipparcos results to demonstrate this from plots of proper motions for
individual stars in the way shown in Figs 2, 3 and 4. Note that the three
stars which lie conspicuous above the others in Figs 2 and 3 are
$\alpha$~UMi, $\delta$~Cep and RT~Aur. These stars are sufficiently close to
the sun that their peculiar velocities have a large effect on their proper
motions.  Their deviations from the mean curve can be accounted for by quite
modest peculiar velocities (10 to 20 $\rm km\ s^{-1}$). In view of the
weighting system adopted, they have little effect on the derived galactic
rotation parameters.

\section{Galactic Rotation Including Higher Order Terms. I}
 In this and the next section we discuss the results obtained when higher
order terms are introduced into the Galactic rotation formulae. The present
section adopts the formulation used by PMB in their analysis of Cepheid
radial velocities. This involves retaining terms in $d/R_{\rm o}$ and also
higher derivatives of $\Theta$. Thus to the level of approximation adopted
by PMB the relation in Galactic longitude becomes;
 \begin{eqnarray}
\kappa \mu_{\ell *} & = (u_{\rm o} \sin \ell - v_{\rm o} \cos \ell)/d - 
\Omega _{\rm o} 
\cos b + \left[ \frac{R_{\rm o}}{d} \cos \ell -\cos b \right] \nonumber \\
&  \times \left[ \frac{2AR_{\rm o}}{R} (1 - R/R_{\rm o}) +
\frac{(R - R_{\rm o})^{2}}{2R} \Theta _{\rm o} ''  +
\frac{(R - R_{\rm o})^{3}}{6R} \Theta _{\rm o} ''' \right] 
\end{eqnarray}
where, $\Theta '' = d^{2} \Theta/dR^{2}$ and $\Theta ''' = d^{3} 
\Theta/dR^{3}.$

We have not attempted to derive a value of $R_{\rm o}$ from the proper
motions but have adopted a value from PMB either directly or adjusted to the
Hipparcos parallax distance scale (section~3) by using fig 5 of PMB.

Table~5 gives the results of various analyses. Solution~1 is the radial
velocity solution adopted by PMB. It should be noticed that the errors given
in this case are ``internal'' errors and do not include any uncertainty in
the distance scale adopted by PMB. For instance if their PLC zero-point is
uncertain by 0.1 mag then the rms error of their derived value of $A$ is $\rm
0.7\ km\ s^{-1}kpc^{-1}$ due to this cause alone. Solution~2 shows the effect
on some of these constants of increasing all the PMB distance moduli by 0.2
mag. These results were derived from fig 5 of PMB. The remaining solutions
are from the proper motion data. For solutions 3 to 13 the adopted distances
were derived from the PL relation obtained from the Hipparcos
trigonometrical parallaxes in paper~1 together with the PC relation adopted
there.  In solutions 14 to 19 the distances were  determined from the PLC
relation derived in section~3 above together with the reddenings discussed
there.

In several of the proper motion solutions the values of $u_{\rm o}$ and
$v_{\rm o}$ were fixed at the values derived by PMB (solution~1).  Similarly
in some solutions the values of $R_{\rm o}$, $\Theta _{\rm o}''$ and
$\Theta_{\rm o}'''$ were fixed as the values from solutions 1 or 2, or
$\Theta_{\rm o}''$ and $\Theta _{\rm o}'''$ were taken as zero. Solution~3
in which only $R_{\rm o}$ was fixed yields values of $\Theta_{\rm o}''$ and
$\Theta_{\rm o}'''$ which are of opposite sign to the values obtained from
the radial velocities. However, the value of $v_{\rm o}$ which goes with
this solution is anomalous (much larger than expected for young objects). If
the local solar motion is fixed at the PMB value (solution~4) the values of
$ \Theta _{\rm o}''$ and $\Theta _{\rm o}'''$ do not differ significantly
from zero.  When the values of $R_{\rm o}$, $\Theta_{\rm o}''$ and
$\Theta_{\rm o}'''$ are fixed at the various values discussed above the
values of the local solar motion derived are close to expected values for
young objects (see Table~7 section~9), e.g.\ solutions 5, 7 and 9. The value
of $A$ obtained (solution~11) when a flat rotation curve is assumed is
significantly lower than other values. This is discussed further in section
7. The value of $\Omega _{\rm o}$ is evidently rather insensitive to the
precise method of solution.

To the extent that the proper motion results have failed to confirm the
values of $\Theta_{\rm o}''$ and $\Theta_{\rm o}'''$ derived from radial
velocities by PMB, there must be a question as to whether the values for
these quantities given by PMB are of general significance for the Galaxy or
simply the effects of random or group motions (including streaming along
spiral arms) of the sample of objects they studied. Nevertheless, in
comparing the values of $A$ determined from proper motions and from radial
velocities it seemed best to fix the values of $u_{\rm o}$, $v_{\rm o}$,
$R_{\rm o}$, $\Theta_{\rm o}''$ and $\Theta_{\rm o}'''$ from the radial
velocity solutions, although setting the last two quantities to zero yields
negligibly different results. For our final solution (solution~19 of
Table~5) we adopt $R_{\rm o}=8.5\ {\rm kpc}$ (see below) and corresponding
values of $\Theta_{\rm o}''$ and ~$\Theta_{\rm o}'''$ from fig 5 of PMB
together with the values of $u_{\rm o}$ and $v_{\rm o}$ from PMB.

Thus we adopt from the proper motions\\
$ A = 14.82 \pm 0.84\ \rm km\ s^{-1}$, $ B = -12.37 \pm 0.64\ \rm km\ s^{-1}$
and $\Omega _{\rm o} = 27.19 \pm 0.87\ \rm km\ s^{-1}$.\\
These results are essentially independent of the adopted distance scale
or of scaling the PMB values of $R_{\rm o}, \Theta''_{\rm o}$ and
$\Theta'''_{\rm o}$ (compare solutions 15, 17 and 19 of table 5).
Comparison of this adopted value of $A$ with the value from radial velocities
($15.92 \pm 0.34$) leads (via fig 5 of PMB) to a revised zero-point
of the PLC relation (equation 3) of \\
$\rho _{2} = -2.43 \pm 0.13$\\
 However, the reddenings adopted by PMB differ slightly in the mean from
those in Table~1 and this must be taken into account in comparing the value
of $\rho_{2}$ derived from proper motions and radial velocities with that
derived in section~3 directly from the trigonometrical parallaxes. The
reddening difference is $\Delta E_{B-V} = 0.011 \pm 0.003$ (from 194
Cepheids in common) with the reddenings of PMB being greater. Taking this
difference into account we obtain;\\
 $\rho_{2} = -2.42 \pm 0.13$\\
which agrees closely with the value derived from the
trigonometrical parallaxes, viz;\\
$\rho_2 = -2.38 \pm 0.10$.

Whilst this forms a very useful check on the distance scale derived from the
Hipparcos trigonometrical parallaxes, it should be realised that any
statistical parallax determination is dependent on the model adopted for
galactic motions and therefore does not have the fundamental status of the
trigonometrical parallaxes. 

PMB derive $R_{\rm o} = 8.09 \pm 0.30\ \rm kpc$ from radial velocities,
where the quoted error does not take into account uncertainties in their
adopted distance scale. This may be revised using the Hipparcos
trigonometrical parallax zero-point and accounting for the difference in
reddening systems. One then obtains (using fig 5 of PMB),\\
 $R_{\rm o} = 8.5 \pm 0.5$ kpc,\\ 
where the standard error now takes into account the uncertainty in the PLC 
zero-point.

\section{Galactic Rotation Including Higher Order Terms. II}
 In a recent paper Metzger et al.\ (1997 =MCS) have re-investigated the
galactic kinematics of Cepheids from radial velocities and have in
particular found some evidence for a weak ellipticity of the galactic disc.
The zero-point of their adopted PL relation was re-evaluated in section~3
above. In the approximation adopted by MCS, which includes the assumption
that the rotational velocity ($\Theta$) is constant, so that, $\Theta _{\rm
o} = 2AR_{\rm o}$, the proper motions in galactic longitude should be given
by the following expression
 \begin{eqnarray}
\kappa \mu _{\ell *}& =  (u_{\rm o} \sin \ell - v_{\rm o} \cos \ell)/d \nonumber \\
& - \Omega _{\rm o}[(1-R_{\rm o}/R)R_{\rm o} \cos \ell /d + R_{\rm o} \cos b/R]\nonumber \\ 
& +  f[R_{\rm o} \sin (2\phi + l)/dR - \sin 2\phi \cos b/R - \sin \ell /d]
\end{eqnarray}
where
\begin{equation}
\sin \phi = d \cos b \sin \ell /R
\end{equation}
 and $f$ is a constant equal to $\Theta _{\rm o} s (R_{\rm o})$. $s(R_{\rm
o})$ is the galactic ellipticity term as defined by MCS. The radial velocity
solutions of MCS lead to $f = +10.2 \pm 3.8\ \rm km\ s^{-1}$.  A variety of
solutions have been carried out using the Hipparcos proper motions to solve
equation 13. These include adopting values for $u_{\rm o}$ and $v_{\rm o}$
from the radial velocity solutions of PMB (see previous section) and MCS
(9.30 and 13.50 $\rm km\ s^{-1}$).  The results are shown in Table~6 where
the first solution is the radial velocity solution adopted by MCS.  In
carrying out out these analyses distances have been determined using the PL
and PC relations discussed in paper~1. As indicated above the reddening
corrections adopted by MCS should be on a system close to this. In no case
did the proper motion solutions yield a value for $f$ which was
significantly different from zero. The errors of $f$ from proper motions are
comparable with those in the MCS radial velocity work. This leaves open the
possibility that the positive value of $f$ found by MCS might reflect group
motions, including flows along spiral arms, rather than a general
ellipticity of the galactic disc. It is worth noticing that in a recent
paper on Cepheid radial velocities Pont et al. (1997) find no significant
evidence for non-axisymmetric components in the rotation of the outer
galactic disc.

As in the case of the solutions of the previous section the
value of $\Omega_{\rm o}$ is quite insensitive to variations in the various
other parameters of the solutions. If we adopt solution~6 in which $R_{\rm
o}$, $u_{\rm o}$, $v_{\rm o}$ and $f$ are fixed at the values adopted by MCS
then $\Omega _{\rm o} = 26.80 \pm 0.83\ \rm km\ s^{-1}kpc^{-1}$ and thus $A =
13.40 \pm 0.42\ \rm km\ s^{-1}kpc^{-1}$. 

The value of $A$ derived from radial velocities scales inversely with the
distance scale (or very nearly so). Thus the value derived by MCS from
radial velocities ($15.47 \pm 1.2$, note that this error is from the two
errors quoted by MCS in quadrature) together with the value just derived
from proper motions, indicates that the MCS scale is underestimated by $0.31
\pm 0.19$ mag. Whilst this estimate agrees within the errors with that
derived from the parallaxes in section~3 ($0.15 \pm 0.09$ mag) it cannot be
regarded with particular confidence. Reference to sections 5 and 6 shows
that when the rotation curve is assumed flat the value of $A$ derived from
the proper motions is smaller than when the rotation curve is not so
constrained.  This clearly illustrates that the model adopted in deriving
distance scales from proper motions and radial velocities may have a
significant effect on the derived result. In this particular case the
problem probably arises because the assumption of a flat rotation curve only
affects second and higher order terms in the radial velocity solutions, but
affects the proper motion solutions in the first order terms.

MCS derive a value of $R_{\rm o} = 7.66 \pm 0.54\ \rm kpc$ (combining their
two quoted error estimates). The quoted error allows for uncertainty in
their adopted distance scale. The PL zero-point (section~3) from the
trigonometrical parallaxes scales this result to $8.2 \pm 0.6$ kpc. As MCS
point out the value of $R_{\rm o}$ that they derive is sensitive to the
galactic ellipticity term. If this is set equal to zero a further increase
in $R_{\rm o}$ of about 4 percent is necessary which would bring their value
to about 8.5 kpc, i.e.\ to the value derived above from the PMB work.

\section{An Estimate of $\Theta_{\rm o}'$ and the Distribution of 
Velocity Residuals}

 Tables 4, 5, and 6 show that, apart from the solutions in which
$\Omega _{\rm o}$ was constrained to equal $2A$, the former is always
less than the latter. Since,
\begin{equation}
\Theta_{\rm o}' = (d\Theta/dR)_{\rm o}  = \Omega _{\rm o} - 2A
\end{equation}
this indicates that in the mean the rotation curve is declining over the
region covered by the Cepheid proper motions. Taking into account the
covariance of $\Omega _{\rm o}$ and $A$ in the various solutions the values
of $\Theta'_{\rm o}$ are $\rm -2.8 \pm 1.3\ km\ s^{-1} kpc^{-1}$ for
the first order solution (Table~4 solution~2) and $-2.4 \pm 1.2$ for
solution~19 of Table~5. This is thus a 2-$\sigma$ result. Evidence for a
declining rotation curve at the Sun's distance from the Galactic Centre is
strengthened if we accept the radial velocity result (PMB, see section~6
above) that $\Theta_{\rm o}''$ is also negative.                            

As Fig 1 shows, the Cepheids in the Hipparcos sample cover a range in $R$
from about 6 kpc to 12 kpc (adopting $R_{\rm o} = 8.5$ kpc).  There is other
evidence for a declining rotation curve over this range of galactocentic
distances. For instance Brand \& Blitz (1993) derived a rotation curve from
radial velocities of HII regions which shows a similar effect. Their fig 5
suggests a somewhat more negative value of $\Theta'_{\rm o}$ than the one
derived from Cepheid proper motions, but in view of the uncertainties the
difference is probably not significant. Brand \& Blitz fit a gradually
rising rotation curve to their complete data set and attribute the declining
portion to the effects of local streaming motions. The fact that the decline
is seen in both the radial velocities and the proper motions which sample
the velocity field in different ways, must strengthen the view that the
decline is a major kinematic feature of this region of the Galaxy. It should
also be noted that Binney \& Dehnen (1997) have argued that the rise seen in
the Brand-Blitz rotation curve for $R > 12$ kpc may be an artifact of the
spatial distribution of objects in the Brand-Blitz sample and that the
rotation curve may in fact continue to decline in the outer regions of the
Galaxy. More recently Pont et al. (1997) have given evidence from radial
velocities of Cepheids in the outer galactic disc that the rotation curve
declines, at least initially, beyond the solar distance from the Centre.

\begin{figure*}
\centering
\epsffile[-7 85 425 340]{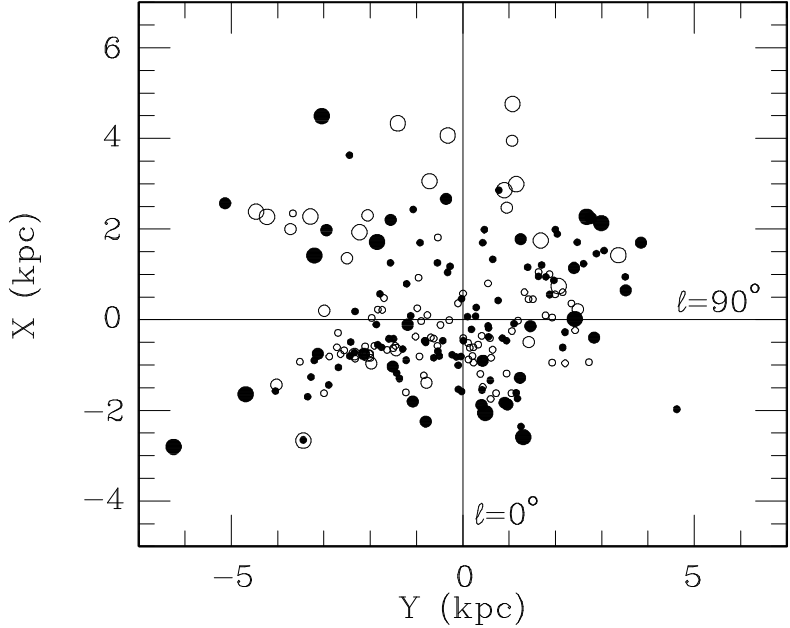}
 \caption{ The velocity residuals derived from solution~19 of Table~5 for
the proper motions in galactic longitude are shown plotted at their
projected positions in the galactic plane. The size of the symbol indicates
the size of the residual without regard to sign. Large circles, $\rm > 40\
km\ s^{-1}$; medium circles, $\rm 20 - 40\ km\ s^{-1}$; small circles, $\rm
0 - 20\ km\ s^{-1}$. Filled circles are for positive residuals and open
circles for negative residuals.}
 \end{figure*} 
 Figure 5 shows the residual velocities of the Cepheids in galactic
longitude as seen projected on the Galactic Plane. These were derived from
the proper motions adopting solution~19 of Table~5. It should be borne in
mind that on the average the uncertainty in these velocity residuals
increases with distance (since they are derived from proper motions). It has
long been known that the radial velocity residuals for young objects (OB
stars, Cepheids) are not randomly distributed but tend to show clumping on a
scale of order of a kiloparsec (Weaver 1964, Feast 1967) which may possibly
be associated with spiral structure (Humphreys 1972). Recent radial velocity
work on Cepheids continues to show evidence of such clumping (PMB, MCS).
There is some slight suggestion of clumping of velocity residuals in Fig 5
(e.g.\ for distant stars in the anti-centre direction) but the evidence is
not strong and there is no very obvious correlation of the velocity
residuals from proper motions with those from radial velocities (e.g.\ fig 7
of PMB).

\section{The Local Solar Motion}
 A large number of determinations of the local solar motion relative to
young objects have been made in the past. No attempt is made to review
them all here. Table~7 contains the results of recent Cepheid work. Solution
1 shows the values of $u_{\rm o}$ and $v_{\rm o}$ derived by PMB from radial
velocities, and solution~2  shows the results of MCS, also from radial
velocities. Solutions 3 and 4 are mean values derived in the present paper
from the proper motions in galactic longitude. Solution~3 is a straight mean
of the results in Table~5 for solutions 5, 7, 9, 12, 14  and 16. Similarly
solution~4 is a straight mean of solutions 2, 4, 7, 9, 10 and  12 in Table~6.
These solutions are given primarily to show that the values of $u_{\rm o}$
and $v_{\rm o}$ derived from the proper motions do not differ significantly
from those obtained from the radial velocities.

The proper motions in galactic latitude can been used to derive a value for
the component of the solar motion relative to Cepheids towards the north
galactic pole, $ w_{\rm o}$. This quantity cannot be satisfactorily
determined from radial velocities since Cepheids and other young objects lie
mainly at low galactic latitude. The equation we have used is,
 \begin{eqnarray}
{w_{\rm o}} & = & -\kappa d \mu_b / \cos b + u_{\rm o}\cos\ell \tan b
+v_{\rm o}\sin \ell \tan b \nonumber \\
 && -\frac{R_{\rm o}}{d}\sin\ell \tan b 
 \left[ 2A (\frac{R_{\rm o}}{R}-1)   \right. \nonumber \\
 & &\left. + \frac{(R-R_{\rm o})^2}{2R}\Theta_{\rm o}^{''}+ 
\frac{(R-R{o})^3}{6R}\Theta_{\rm o}^{'''}\right] .
\end{eqnarray}

Since Cepheids are generally close to the galactic plane all  of the terms
on the right hand side of this equation except the first are small. We have
adopted constants from solution~19 of Table~5 with distances from a PLC
solution as in section~3. Table~8 shows the results obtained with the
material divided up in various ways.  The Cepheids X Cyg ($\rm HIP\,
102276$) and RV Sco ($\rm HIP\, 83059$) have relatively large residuals from
the general solution. Thus solutions omitting them are given. Figure 6 shows
the velocity residuals in galactic latitude from solution~1 as seen
projected on the Galactic Plane. As with the velocity residuals in the Plane
(Fig 5) there is some slight suggestion of a non-random distribution of
residuals. There is also some suggestion that the mean residual might be
different for stars in the anti-centre direction (positive X) and those with
negative X. But the standard errors of solutions 4, 5 and 6 show that any
overall difference is hardly significant.
 \begin{figure*}
\centering
\epsffile[7 85 425 340]{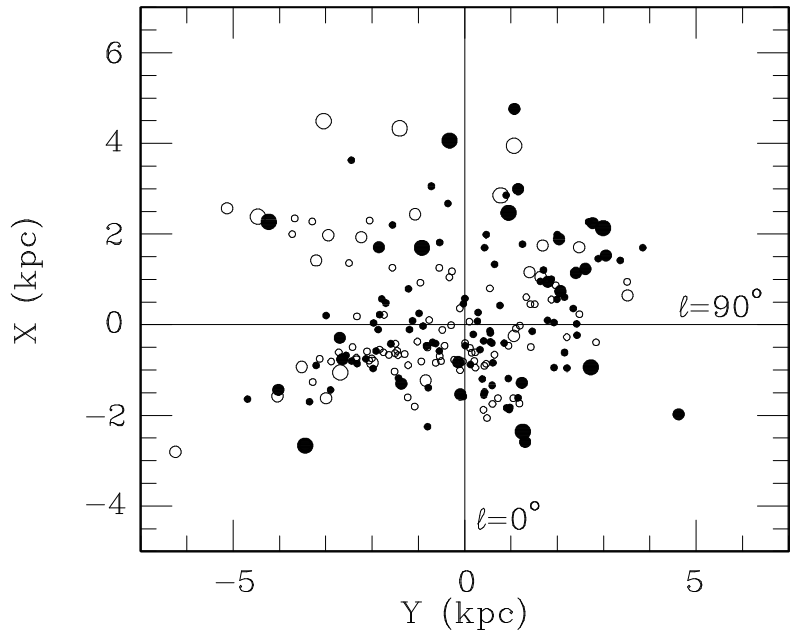}
\caption{The velocity residuals from the proper motion solution in galactic
latitude (solution~1 of Table~8) shown plotted on the galactic plane. 
Symbols as in Fig 5.}
\end{figure*}

Since there may be some concern that the constraints placed on the rotation
curve in the MCS type analyses may affect the derived value of $v_{\rm o}$,
the best values for the components of the local solar motion from Cepheids
are probably the radial velocity values of $u_{\rm o}$ and $v_{\rm o}$ from
solution~1 of Table~7 and the proper motion solution for $w_{\rm o}$
(solution~1 of Table~8). These are given as solution~6 in Table~7. For
comparison the values for young objects derived by Delhaye (1965) are given
as solution~7. It is clear that the modern results remain quite similar to
his values.

\section{Conclusions}
 The following parameters have been derived:\\
$A =  14.82 \pm 0.84\ \rm km\ s^{-1}kpc^{-1}$;\\
$B =  -12.37 \pm 0.64\ \rm km\ s^{-1}kpc^{-1}$;\\
$\Omega_{\rm o} = 27.19 \pm 0.87\ \rm km\ s^{-1}kpc^{-1}$;\\
$(d \Theta/dR)_{\rm o} = -2.4 \pm 1.2\ \rm km\ s^{-1}kpc^{-1}$;\\
$u_{o} =+ 9.3\ \rm km\ s^{-1}$ (adopted from radial velocity solutions);\\
$v_{o} =+ 11.2\ \rm km\ s^{-1}$ (adopted from radial velocity solutions);\\
$w_{o} =+ 7.61 \pm 0.64\ \rm km\ s^{-1}$;\\
$R_{\rm o} = 8.5 \pm 0.5\ \rm kpc$.\\
 The distance scale implied by the comparison of the value of $A$ derived
from the proper motions with that derived from radial velocities is found to
be in agreement with that derived directly from the Hipparcos
trigonometrical parallaxes of the same stars.

\subsection*{Acknowledgements}
 We are grateful to the Hipparcos team for early access to this remarkable
data set, to Metzger et al.\ for their preprint and to Dr F.\ Pont and the
referee (Prof J. Binney) for some helpful comments.

\end{document}


\begin{table*}
\caption{Photometric Data}
\begin{tabular}{@{}rrcrcrc}
\\
\multicolumn{1}{c}{HIP} & \multicolumn{1}{c}{$\langle V \rangle$} & 
\multicolumn{1}{c}{$\langle B\rangle -\langle V \rangle$} &
\multicolumn{1}{c}{$E_{B-V}$} &   
\multicolumn{1}{c}{$\rm \log P$} & \multicolumn{1}{c}{GCVS}&
\multicolumn{1}{c}{Note}\\
\multicolumn{1}{c}{ID} & \multicolumn{3}{c}{(mag)} & & 
\multicolumn{1}{c}{name}\\
        1162&  9.127 &   0.989 &   0.345 &   0.764 & FM Cas \\
        1213&  9.868 &   0.992 &   0.464 &   0.610 & SY Cas \\
        2085&  7.733 &   0.582 &   0.115 &   0.330 & TU Cas \\
        2347&  8.969 &   1.154 &   0.490 &   0.903 & DL Cas \\
        3886&  9.935 &   1.147 &   0.430 &   0.653 & XY Cas \\
        5138& 10.697 &   1.245 &   0.475 &   0.778 & VW Cas \\
        5658& 11.338 &   1.110 &   0.520 &   0.629 & UZ Cas \\
        5846& 10.920 &   1.550 &   0.948 &   0.797 & BP Cas \\
        7548&  9.117 &   1.096 &   0.382 &   1.170 & RW Cas \\
        8312& 10.366 &   1.309 &   0.796 &   0.662 & BY Cas &o \\
        8614& 10.724 &   1.143 &   0.553 &   0.793 & VV Cas \\
        9928&  9.312 &   1.158 &   0.500 &   1.037 & VX Per \\
       10332& 11.664 &   1.027 &   0.538 &   0.660 & UX Per \\
       11420&  9.853 &   1.419 &   0.817 &   1.135 & SZ Cas \\
       11767&  1.982 &   0.598 &  --0.034 &   0.754 & $\alpha$ UMi&o \\
       12817& 10.848 &   1.181 &   0.599 &   0.583 & DF Cas \\
       13367&  5.970 &   0.703 &   0.204 &   0.440 & SU Cas &o\\
       18260&  8.691 &   1.351 &   0.590 &   1.215 & RW Cam \\
       19057&  7.682 &   1.193 &   0.553 &   0.898 & RX Cam \\
       19978& 11.158 &   1.155 &   0.490 &   0.632 & SX Per \\
       20202&  9.723 &   1.302 &   0.676 &   0.697 & AS Per \\
       21517&  6.530 &   0.852 &   0.274 &   0.651 & SZ Tau & o\\
       22275&  7.492 &   1.055 &   0.464 &   0.810 & AW Per \\
       22445&  9.020 &   1.029 &   0.385 &   1.046 & SV Per \\
       23210& 10.455 &   1.218 &   0.593 &   1.012 & AN Aur \\
       23360&  7.655 &   1.009 &   0.246 &   1.065 & RX Aur \\
       24105&  9.427 &   1.062 &   0.446 &   0.903 & BK Aur \\
       24281&  9.074 &   1.000 &   0.407 &   1.006 & SY Aur \\
       24500& 10.332 &   1.375 &   0.565 &   1.260 & YZ Aur \\
       25642&  9.607 &   0.911 &   0.394 &   0.587 &  Y Aur \\
       26069&  3.756 &   0.799 &   0.079 &   0.993 & $\beta$ Dor \\
       27119&  8.217 &   0.847 &   0.355 &   0.606 & ST Tau \\
       28625& 10.007 &   1.025 &   0.566 &   0.743 & RZ Gem \\
       28945&  9.721 &   1.061 &   0.316 &   1.053 & AA Gem \\
       29022& 11.381 &   0.924 &   0.402 &   0.590 & CS Ori \\
       30219&  8.219 &   1.048 &   0.229 &   1.183 & SV Mon \\
       30286&  8.412 &   0.945 &   0.362 &   0.879 & RS Ori \\
       30541&  6.123 &   1.168 &   0.172 &   1.432 &  T Mon \\
       30827&  5.446 &   0.595 &   0.047 &   0.571 & RT Aur \\
       31404&  6.950 &   0.889 &   0.251 &   0.898 &  W Gem \\
       31624& 10.306 &   1.337 &   0.764 &   0.731 & CV Mon \\
       32180&  9.857 &   0.694 &   0.113 &   0.578 & AD Gem \\
       32854& 10.960 &   1.096 &   0.512 &   0.940 & TX Mon \\
       33014& 11.048 &   1.195 &   0.559 &   0.597 & EK Mon \\
       33520& 10.761 &   1.116 &   0.404 &   0.871 & TZ Mon \\

\end{tabular}
\end{table*}
\newpage
\begin{table*}
\begin{tabular}{@{}rrcccrc}
\multicolumn{1}{c}{HIP} & \multicolumn{1}{c}{$\langle V \rangle$} & 
\multicolumn{1}{c}{$\langle B\rangle -\langle V \rangle
$} & \multicolumn{1}{c}{$E_{B-V}$} & 
\multicolumn{1}{c}{$\rm \log P$} & \multicolumn{1}{c}{GCVS}&\multicolumn{1}{c}{Note}\\
\multicolumn{1}{c}{ID} & \multicolumn{3}{c}{(mag)} & & 
\multicolumn{1}{c}{name}\\
       33791& 10.067 &   1.165 &   0.512 &   0.904 & AC Mon \\
       34088&  3.918 &   0.798 &   0.080 &   1.006 & $\zeta$ Gem \\
       34527& 10.582 &   1.175 &   0.577 &   0.669 & TV CMa \\
       35212&  8.110 &   0.847 &   0.236 &   0.670 & RY CMa \\
       35665&  9.697 &   1.004 &   0.456 &   0.629 & RZ CMa \\
       35708&  9.561 &   0.970 &   0.334 &   0.845 & TW CMa \\
       36088&  9.915 &   1.212 &   0.566 &   1.092 & SS CMa \\
       36617& 11.365 &   1.065 &   0.475 &   0.632 & VW Pup \\
       36666&  8.328 &   0.610 &   0.165 &   0.479 & VX Pup \\
       36685&  8.507 &   1.208 &   0.417 &   1.414 &  X Pup \\
       37207&  9.631 &   1.158 &   0.478 &   1.365 & VZ Pup \\
       37511& 10.554 &   0.874 &   0.388 &   0.742 & WW Pup \\
       37515&  9.063 &   0.968 &   0.310 &   0.951 & WX Pup \\
       38063&  9.863 &   1.049 &   0.341 &   1.133 & AD Pup \\
       38907&  7.371 &   0.838 &   0.232 &   0.706 & AP Pup \\
       38944& 10.569 &   0.791 &   0.273 &   0.720 & WY Pup \\
       38965&  8.669 &   1.337 &   0.555 &   1.479 & AQ Pup \\
       39144& 10.326 &   0.789 &   0.203 &   0.701 & WZ Pup \\
       40155&  5.695 &   0.579 &   0.060 &   0.782 & AH Vel& o \\
       40178&  7.957 &   0.783 &   0.223 &   0.824 & AT Pup \\
       40233&  7.028 &   1.434 &   0.484 &   1.617 & RS Pup \\
       41588&  7.375 &   0.870 &   0.191 &   0.826 &  V Car \\
       42257&  7.089 &   1.129 &   0.296 &   1.310 & RZ Vel \\
       42321&  8.032 &   0.934 &   0.284 &   0.667 &  T Vel \\
       42492& 10.017 &   0.991 &   0.510 &   0.495 & AP Vel \\
       42831&  8.121 &   1.151 &   0.356 &   1.370 & SW Vel \\
       42926&  8.277 &   0.885 &   0.276 &   0.980 & SX Vel \\
       42929&  9.704 &   1.195 &   0.520 &   0.768 & ST Vel \\
       44847&  7.635 &   1.175 &   0.450 &   0.840 & BG Vel \\
       45949&  7.589 &   0.788 &   0.149 &   0.641 &  V Vel \\
       46746&  9.520 &   1.518 &   0.682 &   1.049 & DR Vel \\
       47177& 10.262 &   1.243 &   0.646 &   0.853 & AE Vel \\
       47854&  3.735 &   1.260 &   0.194 &   1.551 & {\it l} Car \\
       48663&  9.364 &   1.043 &   0.388 &   0.857 & GX Car \\
       50244& 10.700 &   1.089 &   0.410 &   0.693 & CN Car \\
       50615& 10.261 &   0.979 &   0.443 &   0.619 & GZ Car \\ 
       50655&  8.372 &   1.367 &   0.558 &   1.449 & RY Vel \\
       50722&  8.851 &   0.928 &   0.182 &   0.990 & AQ Car \\
       51142&  9.426 &   0.971 &   0.478 &   0.728 & UW Car \\
       51262&  8.714 &   1.124 &   0.398 &   1.259 & YZ Car \\
       51338&  8.308 &   0.627 &   0.126 &   0.566 & UX Car \\
       51653&  8.084 &   0.585 &   0.126 &   0.561 &  Y Car \\
       51894& 10.654 &   1.162 &   0.530 &   0.844 & XX Vel \\
       51909&  9.323 &   0.875 &   0.206 &   0.716 & UZ Car \\
       52380& 10.318 &   0.854 &   0.352 &   0.459 & EY Car \\
\end{tabular}
\end{table*}
\newpage
\begin{table*}
\begin{tabular}{@{}rrcccrc}
\multicolumn{1}{c}{HIP} & \multicolumn{1}{c}{$\langle V \rangle$} & 
\multicolumn{1}{c}{$\langle B\rangle -\langle V \rangle
$} & \multicolumn{1}{c}{$E_{B-V}$} & 
\multicolumn{1}{c}{$\rm \log P$} & \multicolumn{1}{c}{GCVS}&\multicolumn{1}{c}{Note}\\
\multicolumn{1}{c}{ID} & \multicolumn{3}{c}{(mag)} & & 
\multicolumn{1}{c}{name}\\
       52538&  7.460 &   1.164 &   0.230 &   1.279 & VY Car \\
       52570&  8.524 &   1.054 &   0.407 &   1.149 & SV Vel \\
       52661&  9.089 &   0.887 &   0.344 &   0.687 & SX Car \\
       53083&  9.743 &   0.890 &   0.427 &   0.670 & WW Car \\
       53397&  9.255 &   1.149 &   0.384 &   1.362 & WZ Car \\
       53536&  9.322 &   1.054 &   0.362 &   1.196 & XX Car \\
       53589&  6.281 &   1.178 &   0.277 &   1.588 &  U Car \\
       53593&  9.782 &   0.953 &   0.389 &   0.630 & CY Car \\
       53867& 11.542 &   1.101 &   0.578 &   0.661 & FN Car \\
       53945&  9.295 &   1.214 &   0.420 &   1.095 & XY Car \\
       54101&  8.601 &   1.266 &   0.371 &   1.221 & XZ Car \\
       54543&  6.824 &   0.867 &   0.110 &   0.888 & ER Car \\
       54621&  9.177 &   0.932 &   0.385 &   0.758 & GH Car \\
       54715&  8.097 &   0.990 &   0.279 &   0.877 & IT Car \\
       54862&  8.323 &   0.739 &   0.195 &   0.802 & GI Car & o\\
       54891&  9.661 &   1.121 &   0.337 &   1.030 & FR Car \\
       55726&  8.830 &   1.009 &   0.320 &   0.725 & AY Cen \\
       55736&  8.636 &   0.653 &   0.177 &   0.660 & AZ Cen & o\\
       56176&  8.186 &   0.758 &   0.183 &   0.741 & V419 Cen \\
       56991&  8.765 &   0.740 &   0.255 &   0.523 & UZ Cen \\
       57130& 11.480 &   1.282 &   0.596 &   1.086 & KK Cen \\
       57260&  9.022 &   0.834 &   0.382 &   0.489 & RT Mus \\
       57649& 10.008 &   0.886 &   0.371 &   0.502 & BK Cen \\
       57884&  9.783 &   1.147 &   0.411 &   1.066 & UU Mus \\
       57978& 10.073 &   0.953 &   0.389 &   0.757 & BB Cen & o\\
       59551&  6.127 &   0.831 &   0.268 &   0.985 &  S Mus \\
       59575& 11.051 &   1.279 &   0.677 &   0.806 & AD Cru \\
       59996&  9.796 &   1.752 &   1.051 &   1.109 & SU Cru \\
       60259&  6.566 &   0.922 &   0.178 &   0.828 &  T Cru \\
       60455&  6.766 &   0.772 &   0.167 &   0.765 &  R Cru \\
       61981&  6.298 &   0.757 &   0.132 &   0.876 &  R Mus \\
       62986&  6.600 &   0.761 &   0.167 &   0.671 &  S Cru\\
       63693&  9.966 &   1.172 &   0.574 &   0.646 & V496 Cen \\
       64969&  8.460 &   1.035 &   0.398 &   0.810 & V378 Cen\\
       66189& 10.242 &   1.347 &   0.418 &   1.177 & VW Cen \\
       66383&  9.855 &   1.622 &   0.751 &   1.532 & KN Cen \\
       66696&  7.818 &   0.982 &   0.275 &   1.040 & XX Cen \\
       67566&  7.653 &   0.792 &   0.213 &   0.706 & V381 Cen \\
       70203&  8.753 &   1.191 &   0.419 &   0.976 & V339 Cen \\
       71116&  6.823 &   0.872 &   0.302 &   0.740 &  V Cen \\
       72773&  5.880 &   0.741 &   0.371 &   0.722 & AX Cir \\
       74448&  9.566 &   1.314 &   0.596 &   0.916 & IQ Nor \\
       75018&  6.660 &   0.722 &   0.164 &   0.530 &  R TrA \\
       76918&  9.229 &   1.622 &   0.921 &   1.102 &  U Nor \\
       77913&  9.513 &   1.340 &   0.646 &   1.102 & SY Nor \\
\end{tabular}
\end{table*}
\newpage
\begin{table*}
\begin{tabular}{@{}rrcccrc}
\multicolumn{1}{c}{HIP} & \multicolumn{1}{c}{$\langle V \rangle$} & 
\multicolumn{1}{c}{$\langle B\rangle -\langle V \rangle$} & \multicolumn{1}{c}{$E_{B-V}$} & 
\multicolumn{1}{c}{$\rm \log P$} & \multicolumn{1}{c}{GCVS}&\multicolumn{1}{c}{Note}\\
\multicolumn{1}{c}{ID} & \multicolumn{3}{c}{(mag)} & & 
\multicolumn{1}{c}{name}\\
       78476&  6.397 &   0.752 &   0.092 &   0.801 &  S TrA \\
       78771& 11.670 &   2.000 &   1.194 &   1.033 & TW Nor \\
       78797& 10.027 &   1.287 &   0.552 &   0.792 & RS Nor \\
       78978&  7.878 &   0.608 &   0.104 &   0.410 &  U TrA \\
       79625& 10.411 &   1.273 &   0.684 &   0.538 & GU Nor \\
       79932&  6.426 &   0.945 &   0.196 &   0.989 &  S Nor \\
       82498&  9.810 &   1.936 &   0.886 &   1.458 & KQ Sco \\
       83059&  7.040 &   0.955 &   0.366 &   0.783 & RV Sco \\
       83674&  7.332 &   0.856 &   0.268 &   0.609 & BF Oph \\
       85035&  6.654 &   0.936 &   0.211 &   0.832 & V636 Sco \\
       85701&  7.965 &   0.975 &   0.335 &   0.656 & V482 Sco \\
       87072&  4.549 &   0.739 &   0.240 &   0.846 &  X Sgr \\
       87173&  8.729 &   1.276 &   0.608 &   0.969 & V500 Sco \\
       87345&  8.016 &   1.480 &   0.702 &   1.308 & RY Sco \\
       87495&  6.150 &   1.385 &   0.630 &   1.234 &  Y Oph \\
       88567&  4.668 &   0.746 &   0.119 &   0.881 &  W Sgr \\
       89013& 10.842 &   1.644 &   1.011 &   0.724 & CR Ser \\
       89276&  6.955 &   0.807 &   0.165 &   0.704 & AP Sgr \\
       89596&  8.023 &   1.404 &   0.438 &   1.339 & WZ Sgr \\
       89968&  5.744 &   0.856 &   0.179 &   0.761 &  Y Sgr \\
       90110& 10.549 &   1.457 &   0.822 &   0.818 & AY Sgr \\
       90241&  8.852 &   1.107 &   0.525 &   0.808 & XX Sgr \\
       90791& 10.006 &   1.140 &   0.654 &   0.623 &  X Sct \\
       90836&  6.685 &   1.091 &   0.412 &   0.829 &  U Sgr \\
       91239& 10.131 &   1.182 &   0.664 &   0.490 & EV Sct \\
       91366&  9.628 &   1.539 &   0.765 &   1.015 &  Y Sct \\
       91613& 10.590 &   1.566 &   0.820 &   0.870 & CK Sct \\
       91697&  9.465 &   1.701 &   0.965 &   1.294 & RU Sct \\
       91706& 10.831 &   1.657 &   0.939 &   1.043 & TY Sct \\
       91738& 11.106 &   1.371 &   0.737 &   0.593 & CM Sct \\
       91785&  9.600 &   1.330 &   0.485 &   1.111 &  Z Sct \\
       91867&  8.211 &   0.944 &   0.329 &   0.565 & SS Sct \\
       92013&  7.483 &   0.905 &   0.284 &   0.712 & V350 Sgr \\
       92370&  7.358 &   1.032 &   0.292 &   0.980 & YZ Sgr \\
       92491&  6.932 &   0.985 &   0.295 &   0.822 & BB Sgr \\
       93063& 11.083 &   1.280 &   0.684 &   0.475 & V493 Aql \\
       93124&  5.372 &   0.756 &   0.173 &   0.806 & FF Aql & o \\
       93399&  9.848 &   1.312 &   0.608 &   0.864 & V336 Aql \\
       93681&  8.617 &   1.429 &   0.540 &   1.234 & SZ Aql \\
       93990&  7.141 &   1.292 &   0.444 &   1.138 & TT Aql \\
       94004&  7.751 &   1.146 &   0.395 &   0.833 & V496 Aql \\
       94094&  8.270 &   1.277 &   0.629 &   0.786 & FM Aql \\
       94402&  8.382 &   1.214 &   0.503 &   0.977 & FN Aql \\
       95118& 10.037 &   1.462 &   0.834 &   0.860 & V600 Aql \\
       95820&  6.446 &   1.024 &   0.341 &   0.847 &  U Aql \\
\end{tabular}
\end{table*}
\newpage
\begin{table*}
\begin{tabular}{@{}rrcccrc}
\multicolumn{1}{c}{HIP} & \multicolumn{1}{c}{$\langle V \rangle$} & 
\multicolumn{1}{c}{$\langle B\rangle -\langle V \rangle$} & \multicolumn{1}{c}{$E_{B-V}$} & 
\multicolumn{1}{c}{$\rm \log P_0$} & \multicolumn{1}{c}{GCVS}&\multicolumn{1}{c}{Note}\\
\multicolumn{1}{c}{ID} & \multicolumn{3}{c}{(mag)} & & 
\multicolumn{1}{c}{name}\\
       96458&  7.128 &   1.275 &   0.654 &   0.903 &  U Vul \\
       96596& 10.710 &   0.847 &   0.229 &   0.746 & V924 Cyg \\
       97150&  6.859 &   0.575 &   0.077 &   0.585 & SU Cyg \\
       97717&  7.243 &   1.465 &   0.431 &   1.653 & SV Vul \\
       97804&  3.897 &   0.789 &   0.146 &   0.856 & $\eta$ Aql \\
       98085&  5.622 &   0.805 &   0.092 &   0.923 &  S Sge \\
       98212&  8.849 &   1.389 &   0.709 &   0.801 &  X Vul \\
       98376&  9.924 &   1.266 &   0.662 &   0.893 & GH Cyg \\
       98852&  8.947 &   1.266 &   0.487 &   1.232 & CD Cyg \\
       99276&  9.873 &   1.008 &   0.376 &   0.640 & V402 Cyg \\
       99567&  9.489 &   1.316 &   0.633 &   0.775 & MW Cyg \\
      101393&  9.432 &   1.477 &   0.570 &   1.179 & SZ Cyg \\
      102276&  6.391 &   1.130 &   0.223 &   1.214 &  X Cyg \\
      102949&  5.754 &   0.635 &   0.039 &   0.647 &  T Vul \\
      103433& 10.069 &   1.704 &   0.791 &   1.304 & VX Cyg \\
      103656&  9.511 &   1.784 &   1.151 &   1.168 & TX Cyg \\
      104002&  9.593 &   1.215 &   0.633 &   0.895 & VY Cyg \\
      104185&  5.774 &   0.538 &   0.004 &   0.549 & DT Cyg & o\\
      104564& 10.601 &   1.439 &   0.763 &   0.860 & V459 Cyg \\
      104877&  9.635 &   1.491 &   0.854 &   0.721 & V386 Cyg \\
      105369&  9.086 &   1.036 &   0.455 &   0.516 & V532 Cyg \\
      106754& 10.456 &   1.283 &   0.675 &   0.787 & V538 Cyg \\
      107899&  8.959 &   0.876 &   0.246 &   0.687 & VZ Cyg \\
      108427& 10.590 &   1.668 &   0.682 &   1.252 & CP Cep \\
      108630&  8.883 &   0.949 &   0.304 &   0.727 & BG Lac \\
      109340&  9.146 &   0.731 &   0.170 &   0.636 &  Y Lac \\
      110964& 11.180 &   1.341 &   0.704 &   0.859 & AK Cep \\
      110991&  3.954 &   0.657 &   0.089 &   0.730 & $\delta$ Cep \\
      111972&  8.415 &   1.095 &   0.339 &   1.037 &  Z Lac \\
      112026&  8.848 &   0.885 &   0.287 &   0.807 & RR Lac \\
      112430&  9.656 &   1.396 &   0.720 &   0.795 & CR Cep \\
      112626&  8.936 &   0.873 &   0.286 &   0.698 &  V Lac \\
      112675&  8.407 &   0.901 &   0.296 &   0.736 &  X Lac \\
      114160&  9.705 &   1.081 &   0.457 &   0.736 & SW Cas \\
      115390& 10.973 &   1.650 &   0.939 &   1.179 & CM Cas \\
      115925& 11.641 &   1.738 &   0.986 &   1.158 & CY Cas \\
      116556&  9.932 &   1.490 &   0.875 &   0.799 & RS Cas \\
      116684& 11.112 &   1.475 &   0.884 &   0.699 & DW Cas \\
      117154& 10.738 &   1.449 &   0.818 &   0.892 & CD Cas \\
      117690&  9.927 &   1.384 &   0.649 &   1.084 & RY Cas \\
      118122&  9.876 &   1.188 &   0.464 &   0.992 & DD Cas \\
      118174& 11.136 &   1.174 &   0.523 &   0.688 & CF Cas \\
\end{tabular}

\end{table*}